\documentstyle[12pt,russian]{article}

\author{V.F.Yudanov, O.N.Martyanov\\Boreskov Institute of Catalysis, Russia}

\title{Reproducible magnetic fluctuations of microwave absorption in
zeolites.}

\date{}

\begin{document}

\maketitle{\bf Abstract.}\\

{\footnotesize For polycrystalline specimens of zeolites treated to thermal oxidation
we have revealed and studied the weak spectra of reproducible magnetic 
microwave absorption fluctuations at a frequency of 9.4 GHz. The 
investigation was carried out by the ESR method over the 77-500 K 
temperature range. The fluctuation spectra have features typical of 
mesoscopic phenomena observed previously in ESR spectra of high temperature
superconductors and metals at temperatures below critical. The similar
fluctuation spectra were obtained at 300 K for some oxides, oxidized
silver and gold particles. The experiments performed are, perhaps, the
first observation of the phenomenon at unusually high temperatures which
has been observed only in superconducting systems.}\\
\\
\\

	An investigation of properties of natural and synthetic zeolites 
which belong in aluminosilicates with a geometric rigid set of pores and 
channels is progressing largely because of their catalytic and adsorption 
activity; this latter is of interest for technical applications and yet to be 
explained physically in many aspects [1].

\begin{sloppypar}
	Zeolites exhibit certain unique properties, in particular, after the 
high-temperature (700-1000 K) treatment in $O_2$ atmosphere [2,3]. This
paper reports the results of an experimental investigation of specific 
microwave absorption by zeolites in the magnetic field. We have studied 
polycrystalline specimens of various zeolites, namely, ZSM-5, ZSM-12, and 
mordenite. When preparing the specimens we have previously performed an 
ion-exchange in aqueous solution of $NH_4Cl$ and then the zeolites were kept
in $O_2$ atmosphere at 850 K for 1-2 hours. An ESR-300 3cm-range Bruker
spectrometer was used in our study. The temperature varied within the 77-
500 K range.
\end{sloppypar}

	1. All the specimens registered in the normal fashion exhibit the 
familiar large-intense broad ESR signals caused by the paramagnetic $Fe^{3+}$
ions which usually present as impurities in zeolites. Against a broad 
spectrum background, however, one can register quite weak absorption 
lines in an unusually large range of values for magnetic field - from a value 
close to zero to that attainable in this spectrometer ((8000 Oe). At a single 
magnetic field sweep, these lines are virtually lost to noises and their 
reliable registration requires a digital accumulation of $\sim$ 10$^2$ scans. The
experiment with ZSM-5 zeolite specimens prepared specially (a content of 
the iron impurities was reduced by a factor of more than 10) showed that the
intensity of weak lines is virtually unchanged in this case. It is felt that the 
origin of the unusual absorption weak lines is not associated with the 
presence of the $Fe^{3+}$ impurities and has another nature. Because of this,
the other spectra are given in a more convenient form derived after the 
subtraction procedure of the broad background constituents. As seen in 
Fig.1, the spectra observed consist of the multitude of the lines, rather 
narrow for a solid, with the breadth on the order of few oersted and peculiar 
splittings of $\sim$5-10 Oe. Superficially these spectra have a "noise" character
but a reasonable criteria for their reality is the rigid reproducibility in an 
independent registration series. Hence the comparison of the spectra (a) 
and (b) in Fig.2 shows that the lines observed  are not any instrument 
artifacts and their positions and shapes are reproducible with a high degree 
of accuracy. We have changed the registration conditions, namely, a 
modulation frequency and amplitude (within the line breadth), the 
integration time constant and the rate of the field sweep breadth. All these 
variations leave the spectrum unchanged and allow the absorption pattern 
to be registered invariable over indefinite periods.

	2. A time-constant reproducibility of spectra is observed, 
however, only with a fixed position of the specimen in the spectrometer 
resonator. The rotation of the specimens even through hundredth fractions 
of the degree causes well-noticeable changes in the spectrum. As seen 
from Fig.3, the large rotation angles result in virtually uncorrelated spectra 
patterns. The reverse rotation fails to restore the absorption pattern to its 
original form. Similarly, if the specimen is remounted on the resonator, a 
general view of the spectrum preserves while one can observe variations in 
the position and shapes of the individual lines in any spectrum section. It is  
obvious that for the polycrystalline zeolite specimen (with the mass about 
0.1 g and specific particle sizes of 1-5 mcm) these phenomena cannot be 
attitutable to the nontotal averaging the interactions responsible for the 
spectra formation. In the absence of the separated direction, the ESR 
spectra of the polycrystalline specimens exhibit only the accumulation 
envelope.

	3. The spectra under study are differently behaved when 
registered in the significantly ingomogeneous external magnetic field. The 
ingomogenity of the magnetic field causes no the individual absorption line 
breadth, opposed to the ESR case. Figure 4 shows the absorption spectra 
of the specimen studied for the homogeneous (a) and inhomogeneous (b) 
magnetic field. For comparison it illustrates also the ESR spectra of the  
nitroxyl radicals registered in ampoules of the same diameter (3 mm) under 
the same conditions. The superfine structure of the spectrum of the 
radicals, specified by the isotropic constant of the interaction with the $^{14}N$
nucleus (I=1, $a_{iso}$=14.7 Oe) is almost entirely blurred with the artificially-
made inhomogeneous magnetic field (the H gradient of the order of 100 Oe/cm).
The spectra resolution of zeolites is free of the essential changes yet.

	4. If the specimen mounted in the spectrometer resonator is 
irradiated with the ultraviolet light during the registration, its spectrum 
changes totally with a slight increasing the intensity. After ceasing 
irradiation the spectrum recovers its original shape. At 77 K, the relaxation 
occurs rather slowly and is easily registered although the spectrum 
accumulation time is 3-5 min. The long-wave effect boundary is 350-400 nm.

	5. The irreversible spectrum changes result from rather quick 
((1000 Oe/s) magnetic field changes (between two registrations) as well as
from the influence of the electrostatic field (the intensity (100 V/cm).
	6. As the temperature of the spectrum registration changes by 
1-2 K, its shape changes entirely. The line intensity falls smoothly (no more 
than 5-7 times) in the 77-450 K range. On further temperature increasing, 
the spectrum intensity falls significantly quicker and the spectrum 
accumulation time (with a reasonable signal/noise ratio) increases abruptly 
leading to problems in the correct evaluation of the spectrum 
disappearance temperature. The last reliable spectrum was registered by 
us at 500 K.

	7. The compounds which provoke the appearance of such 
spectra are chemically stable. Thus the specimens hold their properties 
more than a year on keeping in air at room temperature. The annealing of 
the specimens at about 600 K in $H_2$ atmosphere leads to the
disappearance of the spectra.

	The microwave absorption properties revealed testify to the fact 
that the registered lines are not associated with the usual resonance 
absorption in zeolites. The spectra properties observed, however, are very 
typical of mesoscopic effects in objects of sizes comparable to the electron 
coherence length [4-6]. Well-studied quantum fluctuations of 
superconductivity in the magnetic field at low temperatures are 
reproducible (as for long as is wished) upon the magnetic field scanning 
but each change in the specimen position results in a new (reproducible) 
realization of the random process [7]. These fluctuations (or regular 
oscillations for simple systems) attributable to the quantum interference of 
the charges which follow the different paths without a phase malfunction 
have a periodicity multiple to a flux quantum conducting circuit area. It is 
believed that the noise-type magnetic-field dependencies of the UHF 
absorption are due to the magnetic fluctuations in the conductivity of 
peculiar structures which form in zeolites on the thermal oxidation. Based 
on the specific splitting ($\Delta$H$\sim$5-10 Oe) one can estimate a value order of the
effective circuit size using the relation $\Delta$H=$\Phi _0$/$<l>$ ($\Phi _0$=hc/e$\simeq$4$\cdot$10$^{-7}$ Oe$\cdot$cm$_2$)
to give: $<l>\sim$2$\cdot$10$^{-4}$ cm. This value is close to a mean size of some crystals
in zeolite specimens. Even with this coherence length (abnormally large for 
high temperatures), required for the mesoscopic representation of the 
phenomenon observed, this possibility should not be ignored because there 
are the examples of the ESR absorption spectra with similar characteristics 
and the unique combination of the reproducibility-variability properties. 
Such fluctuation absorption spectra (but of much higher intensity) were first 
registered in high temperature superconducting ceramics (HTSC) [8-10]. In 
the first observations, these spectra were interpreted as a noise increase of 
2-3 orders of magnitude attendant on the ESR spectra registration at 
temperatures below a critical one [8]. The paper [9] shows, however, the 
reproducibility of these "noises" and a change in the spectrum realization 
when a specimen rotates. Fluctuation spectrum, received by us at 77 K for 
the ceramics $YBa_2Cu_3O_{7-x}$ with critical temperature 90 K, pictured on fig.5
by way of illustration. Fluctuations of microwave absorption disappear at 
increase of temperature above critical ( see spectrum (b) on fig.5). The 
authors of [9] have first pointed to the mesoscopic nature of the observed 
phenomena caused by the granular ceramics structure with the system of 
Josephson contacts of superconducting weakly bonded entities. The 
occurrence of the fluctuation absorption spectra appeared also to be 
typical of metals in the superconducting state [11, 12]. A prerequisite to the 
observation of the   fluctuations was the presence of the oxide film on the 
metal particles and the roughness of their surface. The mesoscopic regions 
determined in these works have parameters close to the average sizes of 
HTSC granules or geometric heterogeneities in the case of metal 
superconductors. Because of the unique relation between the fluctuation 
spectra and the presence of the superconducting phases, their 
observations were proposed to use for testing the specimens with a small 
content of the superconducting phase when other known methods are
unsuitable by virtue of the sensitivity reasons [13-14]. It is notable that we 
have failed to reveal an increased diamagnetic contribution in the 
specimens studied when measuring the magnetic susceptibility.
 
	It should be noted that the presence of the structure with the 
mesoscopic microwave absorption fluctuations is not unique for the zeolites 
investigated. By now we have observed such spectra at 300 K for oxidized 
silver and gold particles as well as for $Bi_2O_3$ and $TiO_2$. These results are
suggested to be published elsewhere.
 
	Thus the specimens studied exhibit a certain similarity to 
superconducting oxides both in a preparation technique and very specific 
properties.
 
	We believe that the results obtained are the first direct 
observation of the quantum effects in the charge transport in systems which 
present classic heterogeneous catalysts at temperatures of the real 
chemical processes. Based upon the analogies mentioned above we can 
suggest that the quantum effects in oxidized zeolites and other oxide 
systems are due to the presence of traces of the superconducting phases 
which preserve their properties at temperatures up to 500 K.
 
        The authors thank \fbox {K.I.Zamaraev} , Yu.N.Molin, V.N.Parmon for
interesting discussions and support, A.G.Maryasov, S.P.Moshchenko, 
S.A.Gromilov for their help, and V.I.Romannikov for kindly submitted 
specimens of zeolites. The work was performed by using the equipment of 
the Center of collective use on ESR spectroscopy (Institute of Chemical 
Kinetics and Combustion, Siberian Branch, Russian Academy of Sciences) 
with a financial support of RFFR (Grant 96-03-34020).\\
\\
 
 {\bf References:}\\
        \\
        1. J.A.Rabo ed. "Zeolite Chemistry and Catalysis", ACS 171, 80 (1976).\\
2. H.B.Minachev, D.A.Kondratiev, Uspehi Him. (rus) 52,1921 (1983).\\
3. A.V.Kucherov,A.A.Slinkin,D.A.Kondratyev et.al., J.Mol. Catal. 
37,107 (1986).\\
4. R.A.Webb, S.Washburn, Physics Today 41, 46 (1988).\\
5. B.I.Altshuler, A.G.Aronov, B.Z.Spivak, Letters to JETF (rus) 33, 
101 (1981).\\
6. Y.Geten, Y.Ymry, Ya Azbel, Phys. Rev. Lett. 52, 129 (1984).\\
7. P.A.Lee, A.D.Stone, H.Furuyama, Physical Rev. B 35, 1039,  
(1987).\\
8. J.Stakovich, P.K.Kahol, N.S.Dalas et. al., Physical Rev. B 36, 
7126 (1987).\\
9. V.F.Masterov, €.I.Egorov, N.P.Gerasimov et al., Letters to JETF 
(rus) 46, 289 (1987).\\
10. S.Tyagi, M.Barsoum, K.V.Rao, J.Phys.C 21, L827 (1988).\\
11. K.W.Blazey, A.M.Porties, F.H.Holtzberg, Physica C, 157, 16 
(1989).\\
12. A.S.Kheifets, A.I.Veinger, Physica C 165, 491 (1990).\\
13. B.V.Rozentuller, S.V.Stepanov, M.M.Shabarchina et al., Doklady 
AS USSR (rus) 303, 94 (1988).\\
14. L.S.Lubchenco, S.V.Stepanov, M.L.Lubchenco et al., Journal of 
Phys. Chem. (rus) 66, 2510 (1992).\\

\end{document}